\documentclass[12pt,a4paper]{article}
\usepackage[dvips]{graphicx}
\newcommand{\bea}{\begin{equation}}
\newcommand{\eea}{\end{equation}}
\newcommand{\ber}{\begin{eqnarray}}
\newcommand{\eer}{\end{eqnarray}}

\newcommand{\p}{\partial}

\newcommand{\f}{\frac}

\newcommand{\epsa}{\epsilon}

\begin{document}
\title{A theory for one dimensional asynchronous chemical wave}
\author{A. Bhattacharyay\footnote {Email: arijit@pd.infn.it} \\ Dipartimento di Fisika 'G. Galilei'\\ Universit\'a di Padova\\ Via Marzolo 8, 35131 Padova\\Italy\\}

\date{\today}
\maketitle
\begin{abstract}
We present a theory for experimentally observed phenomenon of one dimensional asynchronous waves. The general principle of coexistence of linear and nonlinear solutions of a dynamical system is underlying the present theoretical work. The result has been proposed analytically and numerical simulations are produced in support of the analytical results.\\  
PACS number(s): 87.10.+e, 47.70.Fw
\end{abstract}
\par
One dimensional asynchronous chemical waves (or 1d spirals) were first reported by J.-J.Perraud et al \cite{per} in 1993. An explanation of this phenomenon had been put forward on the basis of bi-stability of Turing and Hopf states and non-variational effects. Historically, in those days when the above mentioned experiment was done, studying droplet of other states inside a global one under uniform conditions was a topic of tremendous interest. Mainly being influenced by those new observations \cite{ahl,mid,wil,rot} and some numerically simulated asynchronous wave like structures, the bi-stability of Turing and Hopf modes and their interaction was considered to be the basic underlying principle on which a possible explanation of experimentally observed 1d spirals was given. Such localized droplets of global states near a Hopf-Turing instability boundary have been analyzed in many subsequent papers [6-12]. As a result of considering bi-stability the dynamics was considered not to derive form variational principle \cite{per}. Since any other alternative theory is not present, the above mentioned mechanism is till date considered as the basis for such a novel phenomenon of nonlinear waves. In the present letter we are going to demonstrate an alternative scenario for 1D spiral generation. In what follows we will show that sustenance of parity inverted global traveling wave states require the presence of a class of asymmetric localized structures. We show that the mechanism of generation of 1d asynchronous waves is intrinsic to the type of amplitude equation valid close to a Hopf instability boundary i.e. the complex Ginzburg-Landau equation ({\bf CGLE}). Such an amplitude equation is of universal form near the same instability boundary in all systems. As a consequence, the results shown in this letter find its applicability to a broad class of systems including the chemical waves. 
\par

Let us first give a short description of experimental observations reported in ref.\cite{per}. One dimensional spiral pattern was experimentally observed in CIMA reaction. Similar transition between a stationary periodic and a traveling wave like Hopf mode were observed by lowering the starch concentration in the chemical reactor or by keeping a low starch concentration and then increasing the malonic acid concentration. In the latter, it had been observed that when the Hopf state takes up from the Turing state, very often there remain a few spots (considered to be reminiscent of the previous Turing state from the time averaged concentration profile \cite{per}) acting as the source of one dimensional anti synchronous wave trains. Bands of maximum intensity spreads alternatively toward right and left directions with a time delay. These asynchronous wave trains moving in opposite directions are like a section of a 2D spiral by a line passing through its core. It had also been made sure that this 1D spiral is not really a manifestation of an essentially 2D phenomenon in 1D rather it is intrinsically one dimensional in nature. The very basic and primary question that yet has not been answered is exactly who breaks the parity and how. As the experiment and simulations suggest the symmetry breaking agent is local and endogenous in nature, if one identifies the local symmetry breaking agent the next thing is to see how a global asynchronous wave state is supported by this local symmetry breaking.
 
\par
Let us first give an outline of the whole theoretical investigation presented in this letter. The {\bf CGLE} with a cubic nonlinear term is considered so as to keep the system invariant under the change in sign of the order parameter. Being based on the dynamics of {\bf CGLE}, the parity breaking local agent is first analytically identified. Such a class of parity breaking localized instabilities have recently been shown in ref \cite{ari} to destabilize the global nonlinear Turing and Hopf states. Symmetric and asymmetric localized structures are shown to occur on the same footing and thus the present theory accounts for sources or sinks of both parity. An explicit matching of the localized, oscillatory, odd parity structure to global traveling wave solutions of the {\bf CGLE} is shown. The localized structure is setting the boundary condition for the traveling wave states in the inner side and eventually inverts the phase of the traveling waves on opposite directions. To support the analytic findings, numerical evidence of such phenomenon is presented. Numerical result clearly shows the predicted behavior at the core of simulated 1D spiral and away from it.      
               
\par
The {\bf CGLE} has the form of a linear Schroedinger equation when we neglect the nonlinear term. As we know a harmonic well potential localize the solutions in a Gaussian profile, we expect to have the same spatial solution from the linear {\bf CGLE}. We neglect the nonlinear part on the basis that we are  after a localized solution of the form $\f{x}{\sqrt b}e^{-x^2/2b}$. At large $b$ the nonlinear term will be negligible for this localized solution. An important point to be noted is that the Gaussian envelop is not alone responsible for localization. Existence of other global states (which will be shown in the following) are also responsible to restrict the spread of the above mentioned structure close to origin. Thus, close to the origin and for large $b$ the nonlinear term of {\bf CGLE} can safely be dropped when the solution is of the above mentioned form is assumed. The {\bf CGLE} without nonlinear term and for a solution $He^{i\omega t}$ can be written in the form      
\ber\nonumber
i\omega H +\f{\p H}{\p t} &=& \epsa H + (D_r+iD_i)\f{{\p}^2 H}{{\p x}^2}, \\\nonumber
\eer
where $H$ is the slow and large scale dependent amplitude of the Hopf mode which oscillates with a frequency $\omega $ and $\epsa $ is the bifurcation parameter.

Let us take that $H$ as  $He^{\f{-x^2}{2b}}$. With this consideration the above equation changes to
\ber\nonumber
(D_r+D_i)\left[\f{{\p}^2 H}{{\p x}^2}e^{\f{-x^2}{2b}}-\f{2x}{b}\f{\p H}{\p x}e^{\f{-x^2}{2b}}+\f{x^2H}{b^2}e^{\f{-x^2}{2b}}-\f{H}{b}e^{\f{-x^2}{2b}}\right]+(\epsa -i\omega)He^{\f{-x^2}{2b}}  =0\\\nonumber
\eer

\par
The Equations coming out by equating real and imaginary parts look like
\ber\nonumber
D_r\left[\f{{\p}^2 H}{{\p x}^2}-\f{2x}{b}\f{\p H}{\p x}+\f{x^2H}{b^2}-\f{H}{b}\right]+\epsa H=0\\\nonumber
\eer
and
\ber\nonumber
D_i\left[\f{{\p}^2 H}{{\p x}^2}-\f{2x}{b}\f{\p H}{\p x}+\f{x^2H}{b^2}-\f{H}{b}\right]-\omega H=0\\\nonumber
\eer
They look almost the same except for the last terms. Now, considering one that comes from the real parts and rearranging the terms we get
\ber\nonumber
D_r\f{{\p}^2 H}{{\p x}^2}-\f{2D_rx}{b}\f{\p H}{\p x}+\f{D_rx^2}{b^2}H+(\epsa -\f{D_r}{b})H=0\\\nonumber\\
\eer
We are looking for a polynomial solution of the dimensionless amplitude $H$. So, let us nondimensionalize the length scale as $z=x/\sqrt{b}$ which is also suggested by the above form of the equation. We also drop the term containing $x^2/b^2$ which is small for large $b$ where $\f{x^2}{b^2}\sim \f{1}{b}$. Actually, our purpose is served with the expected Hermite polynomial solution of order unity since it has odd parity and we need not consider higher order terms. So, we get the equation  
\ber\nonumber
\f{{\p}^2 H}{{\p z}^2}-2z\f{\p H}{\p z}+\f{b}{D_r}(\epsa-\f{D_r}{b})H=0\\
\eer
\par
The linear Eq.2 admits solutions which are Hermite polynomials and a solution of order zero is obtained when $(\epsa-\f{D_r}{b})$ is equal to zero. When $\epsa-\f{D_r}{b}=\f{2D_r}{b}$ it admits a solution same as Hermite polynomial of order unity. Now, we have explicitly got a small amplitude localized asymmetric solution of the {\bf CGLE} which we will show to support two asynchronous traveling wave state on its two sides. One important thing about this linear localized solution is that it can always have an arbitrary constant factor which can also keep it small apart from the requirement of large $b$. So, in a number of different ways one can justify the non-functioning of the nonlinear term for such a localized solution. Before we show the asynchronous wave solutions we consider the equation got from equating the imaginary terms of the {\bf CGLE}. If both the equations are to have the same asymmetric localized spatial solutions we get a selection for the oscillation frequency given by
\bea
\omega=-\f{\epsa D_i}{D_r}
\eea    
\par
Take the full {\bf CGLE} in the form 
\bea
\f{\p H}{\p t} = \epsa H + (D_r+iD_i)\f{{\p}^2 H}{{\p x}^2}-(\beta_r+i\beta_i)|H|^2H.
\eea

This full {\bf CGLE} can have a one parameter family of traveling wave solutions. So, it is quite reasonable to think about a traveling wave state away from such a core structure. Let us take the traveling wave state in the form $He^{i(kx-\omega t)}$. Consider this traveling wave state be present on the positive side of the x-axis having the localized core at the origin. One has to match this traveling wave amplitude to the spatial profile of the core and will get the amplitude of the traveling wave state from such matching. This can ne done keeping in mind that the amplitude of the traveling wave solution is a function of the parameters of the system and in reality such a matching can always happen by tuning the parameters. Consider that the $\omega$ of the traveling wave to be the same as the oscillation frequency of the central asymmetric core.
Now, putting the expression of $\omega$ of the oscillating core in the above expression we get
$$
|H|^2=\f{D_i(\epsa+\epsa_l)}{D_i\beta_r-D_r\beta_i}
$$ 
where $\epsa_l$ is the local value of the bifurcation parameter where the core has formed. Now, we would like to match the wave that moves on positive x-direction with the positive x-part of the localized core at the point $x=C$ $(say)$ to get
  
\bea
H_r=\f{C}{\sqrt{b}}e^{-C^2/2b+ikC}
\eea 
By matching a similar wave moving toward negative x-direction from near the left hand side of the asymmetric localized core we will get its amplitude $H_l$ as
\bea
H_l=-\f{C}{\sqrt{b}}e^{-C^2/2b+ikC} .
\eea
The boundary conditions on the traveling wave states cause a clear phase difference of $\pi$ in the left and the right moving waves due to obvious reasons ($k$ and C are negative for the left moving wave). Apart from that phase difference everything else are identical on the both sides of the origin. This is actually no surprise given the fact that the {\bf CGLE} is invariant under the inversions $H=-H$ and $x=-x$. So, based on this symmetry argument one can imagine that the initial phase difference in those two waves can be retained over large space and time to produce asynchronous waves. The $e^{ikC}$ term in the amplitude ensures that the points $x=\pm C$ are the origin of the traveling wave states and the oppositely moving waves exist beyond them. The {\bf CGLE} being a nonlinear equation superposition of solutions is not possible. The real traveling wave solution is obtained by considering the amplitude equation for the complex conjugate part of the slow amplitude ($\stackrel{*}{H}$) which will exactly be the same as the one considered here except for the sign of the imaginary terms. This is so because of the symmetry of the complex roots at the linear instability threshold. It is clear from the present analysis that the localized structures will exactly be the same for the both forms of the amplitude equations since they have been got by separating the real and imaginary parts of the equation. So, there is no problem in considering traveling wave forms given by real numbers and matching that with the oscillatory localized core in the middle which will eventually flip the global phases on two sides. The more important point here is the stability of such a combination. In what follows we show the stability of such a system numerically.   
\par
To justify the analytic result we simulated the full nonlinear {\bf CGLE}. We employed finite difference scheme (Crank-Nicolson formula) with implicit method of integration. The nonlinearity is tackled by predictor corrector rule. The numerical integration has been performed on a linear lattice of 1001 points with a small ($\pm 0.01$) parity breaking initial seed symmetrically at two points on the both sides of the center of the lattice. We have employed no flux boundary conditions. Prominent asynchronous waves develop and spread from asymmetric oscillatory core at the middle. Fig.1a shows the space time plot of the concentration waves asynchronously moving on opposite sides. It has a central oscillatory core which is of the same nature as an oscillatory Harmite polynomial of order Unity. To reveal the nature of the central core part more clearly, we have plotted the oscillation amplitude of the space time plot shown in Fig.1a on Fig.1b over the same space time region. The central dark line is the indicator of the zero amplitude central part of the Hermite polynomial of order unity and the two almost parallel bright line on the two sides of this central line indicates the slightly higher amplitude of oscillation of the peaks of the central core compared to that of the traveling waves spreading away on the outer sides of these lines. The relevant parameter values are indicated in the figure caption and it has been seen that similar structures also appear at different parameter regimes. Fig.2a and 2b show such structures at different parameter values. Fig.3 shows the concentration profile at the last time step of what shown in Fig.2a and2b. 
\par
So, we conclude by saying that an asymmetric localized oscillatory structure has been identified as the parity breaking agent for asynchronous one dimensional chemical waves and an alternative theory has been put forward. The present theory rests on the principle of matching and stabilizing localized linear solutions by nonlinear global states of a general nonlinear model in an extended system. Such a situation is very general to our belief and can explain a lot of complex phenomenon as observed in extended nonlinear systems. The linear part with its characteristics symmetry and parity can interact with the nonlinear solutions to generate a range of complexities. The present theory also accounts for the presence of synchronous waves originating from symmetric sources and sinks as the asymmetric ones on the same footing. Even, the consideration of higher order localized structures is possible and it would be interesting to investigate experimentally and numerically the role played by them in an extended system.

\newpage
{\bf Figure Caption}\\

{{\bf Fig.1a}: The concentration profile of the asynchronous wave generated from a an oscillatory localized structure. The parameter 
values for the upper graph are $d_r=1.0,d_i=0.1,\beta_r=0.01,\beta_i=-0.01$ 
and $\epsa=0.1$.}\\

{{\bf Fig.1b}: The space-time plot of the absolute amplitude of the profiles shown in Fig.1a (parameter values are the same).}\\

{{\bf Fig.2a}: The concentration profile of the asynchronous wave. The parameter 
values for the upper graph are $d_r=0.1,d_i=1.0,\beta_r=0.01,\beta_i=0.1$ 
and $\epsa=0.1$.}\\

{{\bf Fig.2b}: The concentration profile of the asynchronous wave. The parameter 
values for the upper graph are $d_r=0.1,d_i=1.0,\beta_r=0.02,\beta_i=0.1$ 
and $\epsa=0.1$.}\\

{{\bf Fig.3}: The concentration profile of the asynchronous wave at time 300 unit. The parameter values for the upper graph are $d_r=0.1,d_i=1.0,\beta_r=0.01,\beta_i=0.1$ and $\epsa=0.1$. For the lower part $\beta_r=0.02$ and all others are the same as the upper one. }\\

\end{document}